\documentclass[9pt,twocolumn,twoside]{pnas-new}
\usepackage{braket}
\pdfoutput=1 % Guarantees pedflatex processing for arXiv
\usepackage[utf8]{inputenc} % allow utf-8 input
\usepackage[T1]{fontenc}    % use 8-bit T1 fonts
\usepackage{hyperref}       % hyperlinks
\usepackage{url}            % simple URL typesetting
\usepackage{booktabs}       % professional-quality tables
\usepackage{amsfonts}       % blackboard math symbols
\usepackage{nicefrac}       % compact symbols for 1/2, etc.
\usepackage{microtype}      % microtypography % [nopatch=eqnum] leads to options clash
\usepackage{xcolor}         % colors
\usepackage{bm}
\usepackage{graphicx}
\usepackage{xr}             % To reference to supplement
\title{A Waddington landscape for prototype learning in generalized Hopfield networks }

\author[a,b]{ Nacer Eddine Boukacem}
\author[a,c]{ Allen Leary} 
\author[a,d]{Robin Thériault}
\author[a]{Felix Gottlieb}
\author[e,f]{  Madhav Mani}
\author[b,g]{Paul Fran\c cois\textsuperscript{1}}

\affil[a]{Rutherford Physics Building, McGill University, Montr\'eal, QC, Canada}
\affil[b]{D\'epartement de Biochimie et Medecine Mol\'eculaire, Universit\'e de Montr\'eal, Montr\'eal, QC, Canada}
\affil[c]{Regeneron Pharmaceuticals,777 Old Saw Mill River Rd, Tarrytown, NY 10591, USA}
\affil[d]{Scuola Normale Superiore di Pisa, Piazza dei Cavalieri, 7 - 56126 Pisa. Italy}
\affil[e]{Engineering Sciences and Applied Mathematics, Northwestern University, Evanston, IL, USA}
\affil[f]{NSF-Simons Center for Quantitative Biology, Northwestern University, Evanston, IL, USA}
\affil[g]{MILA Qu\'ebec, Montr\'eal, QC, Canada}
\affil[1]{To whom correspondence should be addressed. E-mail: paul.francois@umontreal.ca}

\leadauthor{Boukacem} 

\correspondingauthor{\textsuperscript{1} To whom correspondence should be addressed. E-mail: paul.francois@umontreal.ca}


\begin{abstract}
Networks in machine learning offer examples of complex high-dimensional dynamical systems reminiscent of biological systems. Here, we study the learning dynamics of Generalized Hopfield networks,  which permit a visualization of internal memories. These networks have been shown to proceed through a 'feature-to-prototype' transition, as the strength of network nonlinearity is increased, wherein the learned, or terminal, states of internal memories transition from mixed to pure states. Focusing on the prototype learning dynamics of the internal memories we observe a strong resemblance to the canalized, or low-dimensional, dynamics of cells as they differentiate within a Waddingtonian landscape. Dynamically, we demonstrate that learning in a Generalized Hopfield Network proceeds through sequential 'splits' in memory space. Furthermore, order of splitting is interpretable and reproducible. The  dynamics between the splits are canalized in the Waddington sense -- robust to variations in detailed aspects of the system. In attempting to make the analogy a rigorous equivalence, we study smaller subsystems that exhibit similar properties to the full system. We combine analytical calculations with numerical simulations to study the dynamical emergence of the feature-to-prototype transition, and the behaviour of splits in the landscape, saddles points, visited during learning. We exhibit regimes where saddles appear and disappear through saddle-node bifurcations, qualitatively changing the distribution of learned memories as the strength of the nonlinearity is varied -- allowing us to systematically investigate the mechanisms that underlie the emergence of Waddingtonian dynamics. Memories can thus differentiate in a predictive and controlled way, revealing new bridges between experimental biology, dynamical systems theory, and machine learning.
\end{abstract}

\begin{document}

\thispagestyle{firststyle}

\ifthenelse{\boolean{shortarticle}}{\ifthenelse{\boolean{singlecolumn}}{\abscontentformatted}{\abscontent}}{}

\maketitle

\dropcap{T}he field of machine learning offers fascinating examples of networks with self-organizing dynamics.  During learning, network parameters change and qualitatively novel regimes appear  \cite{Hopfield1982, Ramsauer2020}. The dynamics through qualitatively distinct learning regimes open up new possibilities for their optimization and, thus, training. Task optimization in very high dimensions is now seen as 'easy'  as the very high dimensionality of the networks generally ensures that at least one eigenvalue is positive at critical points -- there is always a direction in which the loss can be further reduced. This suggests that only saddle points (or saddles) are met during optimization driven by gradient descent  \cite{Goodfellow2014}, so that the learning dynamics consist of transitions between saddles (which can itself be sped up using various standard methods). However, the quantitative and qualitative natures of the saddles themselves might be irreproducible, depending on details of the system. Such a view of learning, with potentially a huge number of (random) critical points, contrasts more coarse-grained observations of the typical learning dynamics observed in neural networks. To this end, in the limit of very high dimensionality, or number of parameters, exact results on the biases and errors of learning dynamics have been derived using tools inspired by random matrix theory and statistical mechanics \cite{Paquette2021,Rocks2022}. In addition, information theory metrics have further shown structured and reproducible coarse-grained dynamics during learning \cite{Tishby,Mao2023}. 

Together, these studies suggest the emergence of universal, and thus, simpler, regimes of learning, echoing what is observed in other self-organizing  dynamical systems, in particular in biological contexts \cite{Matsushita2022,Matsushita2023}. Case in point: the advances in high throughput quantitative data in biology combined with mathematical modeling have established that the dynamics of high-dimensional biological networks can often be captured by low-dimensional representations, at least locally. Examples of such a modeling paradigm can be found within diverse biological contexts, including differentiating cells \cite{Rand2021}, developing systems \cite{Seyboldt2022}, embryonic mechanics \cite{Liu2022}, and for specific brain decisions \cite{Thura2022}. The dimensionality reduction inherent to the above approaches is concordant, perhaps inspired, by a classical proposal by Conrad Waddington in the context of cellular differentiation. Waddington introduced the notion of an 'epigenetic landscape', where the evolution of cellular states is represented by a ball rolling in canalized valleys \cite{Waddington1957}. Such valleys can subsequently split, which in a dynamical systems context are referred to as saddles, accounting for binary commitments, eventually leading to a multiplicity of terminal cellular fates. Despite the qualitative nature of such descriptions, Waddingtons' view leads to multiple predictions: for instance, cellular decisions should typically  be binary (i.e. between at most two fates), and one should observe sequences of  well-defined 'progenitor' or 'stem' states leading to multiple fates. Such predictions have been largely verified experimentally \cite{Graf2009}, thus confirming Waddington's intuition and explaining its broad impact on developmental biology. While the landscape paradigm was  qualitative, more rigorous theories have been proposed over the years. In a pioneering work, 'classical' Hopfield networks themselves have been used to reverse-engineer an epigenetic landscape \cite{Lang2014}, and further revealed a 1D reaction coordinate during cellular reprogramming \cite{Pusuluri2018}. In addition, low-dimensional models of cell differentiation, based on catastrophe theory, have been derived \cite{Rand2021}, leading to predictions and applications in specific systems  \cite{Saez2021}. Despite all this, what remains unclear is \textit{how and why low dimensional dynamics emerge from the complex interacting components that comprise a typical biological system \cite{Husain2020, Matsushita2022}?} Evolution might play a role here: for instance, it has been shown that low dimensional manifolds can naturally emerge from the simulated evolution of toy-models for genetic networks \cite{Furusawa2018}, thus suggesting that the low-dimensionality of cellular decisions are an evolutionary spandrel \cite{Gould1979}. Here, we provide a new perspective, demonstrating that canalized, Waddingtonian-like, dynamics emerge as a phase in particular regimes of parameter space within a Generalized Hopfield Network, an architecture able to capture multiple machine learning frameworks such as large language \cite{Ramsauer2020} or diffusion models \cite{Hoover2023}. 

\begin{figure}
  \centering
  \includegraphics[width=\linewidth]{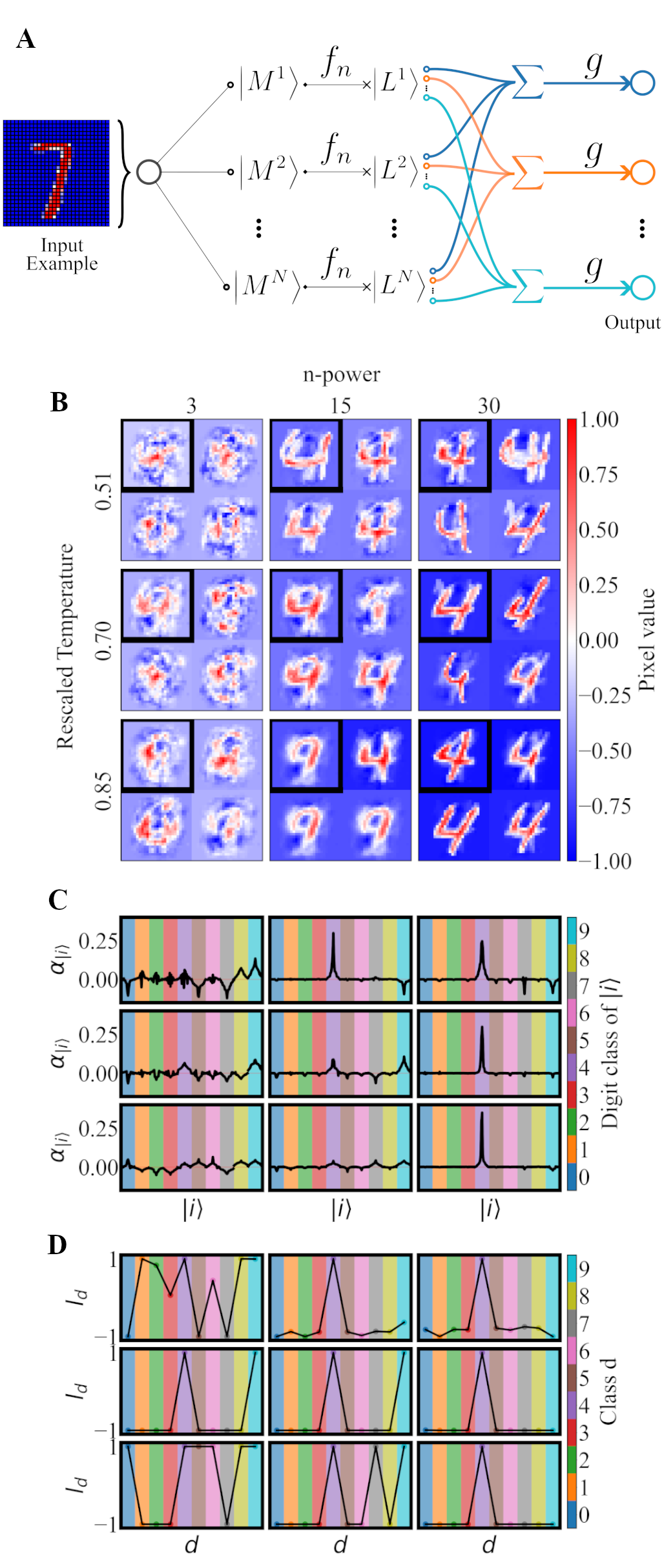}%{Figures (New, Merged)/Figure-0_merged.png}
  \caption{Feature-Prototype transition for generalized Hopfield networks. (A) illustrates the general architecture used. For a given $(n, T_r)$, B) shows samples of the memory final states. The memories chosen are all associated with a final label contributing highly ($\geq 0.99$) to the class $4$. The top left memory for each $(n, T_r)$ is then linearly decomposed using the pseudo-inverse of the training set and in C), the contribution of each training example in that memory is plotted. These linear coefficients are symmetrically re-ordered such that for each class, the maximum is at the center. In D), the label associated to this same memory is shown. For all simulations shown in this figure, the network size was $100$ memories, a training rate of $0.005$ and random gaussian initial conditions. The training set was identical for all runs and consisted of 200 MNIST digits (20 of each class). The simulations vary only in the temperature $T_r=(0.51, 0.7, 0.85)$  and $n=(3, 15, 30)$. }\label{fig:fig0}
\end{figure}

Generalized Hopfield networks were introduced by Krotov and Hopfield in two seminal papers \cite{Krotov2016, Krotov2017}. They represent an elaboration of the classical Hopfield model for associative memory \cite{Hopfield1982}, one of the first modern neural network architectures designed to perform complex tasks. In brief, Hopfield networks rely on a well-designed, spin-glass type, energy function, allowing to  1. store and 2. recover patterns. In this framework, patterns correspond to minima of energy, and can be recovered from noisy initial conditions through gradient descent. In \cite{Krotov2016,Krotov2017} a new hyper-parameter $n$ is added, controlling the strength of nonlinearities of terms in the energy function, thus 'steepening' the energy landscape (other models have also considered an exponential limit, see e.g.  \cite{Demircigil2017}). Krotov and Hopfield further proposed to train a  simple neural network to perform classification using such steepened energy functions, see Fig .\ref{fig:fig0} A for an illustration of the architecture. They consider a model with a hidden layer of (internal) memories (subsequently called $\ket{M^\mu}$, where $\mu$ is the index of a memory considered ). Upon presentation of an input vector $\ket{\sigma}$, the dot products $\braket{M^\mu|\sigma}$ are divided by a temperature $T$ (a hyperparameter of the network), then passed through highly non-linear activation function implementing a steep energy landscape (Rectified Polynomial of order $n$, see in Appendix). Those amplified dot products are then weighted with labels (subsequently called $l^\mu_d$, $d=0 \dots 9$, and gathered together in a label vector $\ket{L^\mu}$) corresponding to different digit identities. Finally, the weighted linear combination of Rectified Polynomials is passed through sigmoidal functions computing categorical scores of the input presented. The intuitive idea here is that the memories encode some patterns, which are recognized in the input then amplified with the help of the $n$ hyper-parameter, eventually allowing for proper digit classification with the help of weighted labels. The parameters of the networks (i.e. the memories and the labels) are trained through gradient descent of a cost function encoding categorical accuracy. Additional technical aspects of the model and training are presented in the Appendix and the Supplement.

Krotov and Hopfield tested their architecture on the classical MNIST dataset  \cite{lecun1998mnist} of handwritten digits. They studied the hidden memories at the end of training, and observed a striking transition in their final states as the hyperparameter $n$ is increased, suggestive of two distinct encoding regimes. They propose that digit classification is transitioning from a 'feature-based' encoding to a 'prototype-based' encoding, Fig. \ref{fig:fig0} B-D.  For low $n$, the internal memories look like overlaps of multiple digits  Fig. \ref{fig:fig0} B top left, suggesting an encoding that is distributed across memories. Conversely, for high $n$, the internal memories look like digits from the training samples, suggesting an encoding based on proximity to exemplar -- or prototype -- digits. Indeed, the number of internal memories used to recognize a specific sample considerably decreases with $n$ \cite{Krotov2016}.  Prototype-based classification is of particular interest to a human observer because it naturally is more understandable/interpretable. It also further prevents undesired properties, e.g. it is more robust to adversarial perturbations specifically designed to fool the classifier \cite{Krotov2017}.

In this work, we further explore the properties of generalized Hopfield networks using a combination of numerical experiments and analytical derivations. We first revisit the results of Krotov and Hopfield, by decomposing memories in a sample basis, in a manner analogous to the Fourier decomposition of complex sounds into pure frequencies, which allows us to quantitatively study multiple aspects of the feature-to-prototype transitions. Going beyond the terminal states of memories, and focusing on the dynamics of learning, we observe two qualitatively different regimes: while for small $n$, the learning dynamics are stochastic and high-dimensional, for higher $n$, they become lower-dimensional, manifesting greater determinism and directionality. In particular, for higher $n$ we typically observe, borrowing Waddingtonian terminology, "canalized" trajectories bringing the learning dynamics proximal to well-defined and reproducible saddles, followed by binary splitting events leading to increased specializations/differentiations of the internal memories.  To better understand this process,  we then study versions of the learning dynamics with a reduced number of memories/samples, where the model can cast analytically with some approximations. We use dynamical systems theory to exhibit at least three different regimes of learning. Generally speaking, our results suggest a surprising diversity in learning dynamics, where characteristics of the learning trajectories influence the final learned outcome that can be predicted in some cases.  We emphasize that the phase of learning dynamics observed in the high $n$ and mid-range of $T$ regime are reminiscent of Waddington's Landscape paradigm for cellular differentiation. This suggests that the dual phenomena of differentiation and canalization are broader than biology, and that the simulated dynamics of learning can be used to model and study the origin and emergence of generic and self-organizing features of complex systems \cite{Goldenfeld2011}.

\section*{Results}

\subsection*{Characterizing feature-to-prototype transitions using Moore-Penrose pseudo-inverse}

 Fig. \ref{fig:fig0}B-D illustrates the feature-to-prototype transition in generalized Hopfield networks, using a small training ensemble of 200 digits, which allows us to better visualize what happens. Because of the learning dynamics (see details in Supplement), reminiscent of quantum mechanics, each memory $\ket{M}$ can be decomposed into a linear sum of samples used in the training set. Formally, we write for each memory (indexed by $\mu$)

\begin{equation}
\ket{M^\mu}=\sum_{\ket{i} \in \mathcal{T}} \alpha^\mu_{\ket{i}} \ket{i}
\end{equation}

where $\mathcal{T}$ defines the entire training set,  $\ket{i}$ is a generic name for a vector in the training set and $\alpha^{\mu}_{\ket{i}}$ a corresponding weight for $\ket{M^\mu}$. Given a memory $\ket{M^\mu}$ and a training set, we can compute such a decomposition using the Moore-Penrose pseudo inverse, which can be  derived from the Singular Value Decomposition of the matrix where line $i$ corresponds to vector $\ket{i}$ (see Supplement). The $\alpha^{\mu}_{\ket{i}}$ are not unique if the vectors in the training set are not linearly independent, which necessarily occurs if the dimension of the  space is smaller than the training set. MNIST has dimension $784=28 \times 28$ (pixels), so that if the training has a smaller number of elements we do not expect any ambiguity with the computation of the $\alpha_{\ket{i}}$. We refer to the Supplemental Figures for illustrations of the reconstruction.

In Fig. \ref{fig:fig0}B we show a sample of memories  $\ket{M^\mu}$ for various $n$ and rescaled temperature $T_r$ (see Appendix) at the end of the training,  such that  $l^\mu_{d=4}=1$, i.e. memories contributing to classify inputs as $4$. We also display corresponding distributions of $\alpha_{\ket{i}}$, ordered as a function of digit categories and ranked symmetrically within a  category in  Fig. \ref{fig:fig0}C, and $l^\mu_d$ as a function of digit category $d$  in Fig. \ref{fig:fig0}D for the same memories. For low $n$, we recover 'feature'-like memories. They clearly consist of positive and negative linear combinations of many different digits with relatively small weights (see e.g. top left corner of  Fig. \ref{fig:fig0} C ), explaining their disordered appearance. Interestingly, these features are not random, e.g. as $n$ is increased,  similar digits  (e.g. typically $4,7,9$, see below) usually contribute significantly positively or negatively. Labels for positive digits typically take the maximum value of $1$ while other labels are $-1$ Fig. \ref{fig:fig0} D .  

As both $n$ and $T_r$ are increased, fewer and fewer input samples contribute to the memories, giving more peaked distributions of  $\alpha_{\ket{i}}$,  Fig. \ref{fig:fig0} C, until one gets a very peaked distribution where very few inputs, associated with the same digit, and correspondingly only one label is present for larger $n,T_r$. This gives rise to well-defined prototypes, e.g. bottom right corner in  \ref{fig:fig0} B .

\begin{figure}
  \centering
  \includegraphics[width=\linewidth]{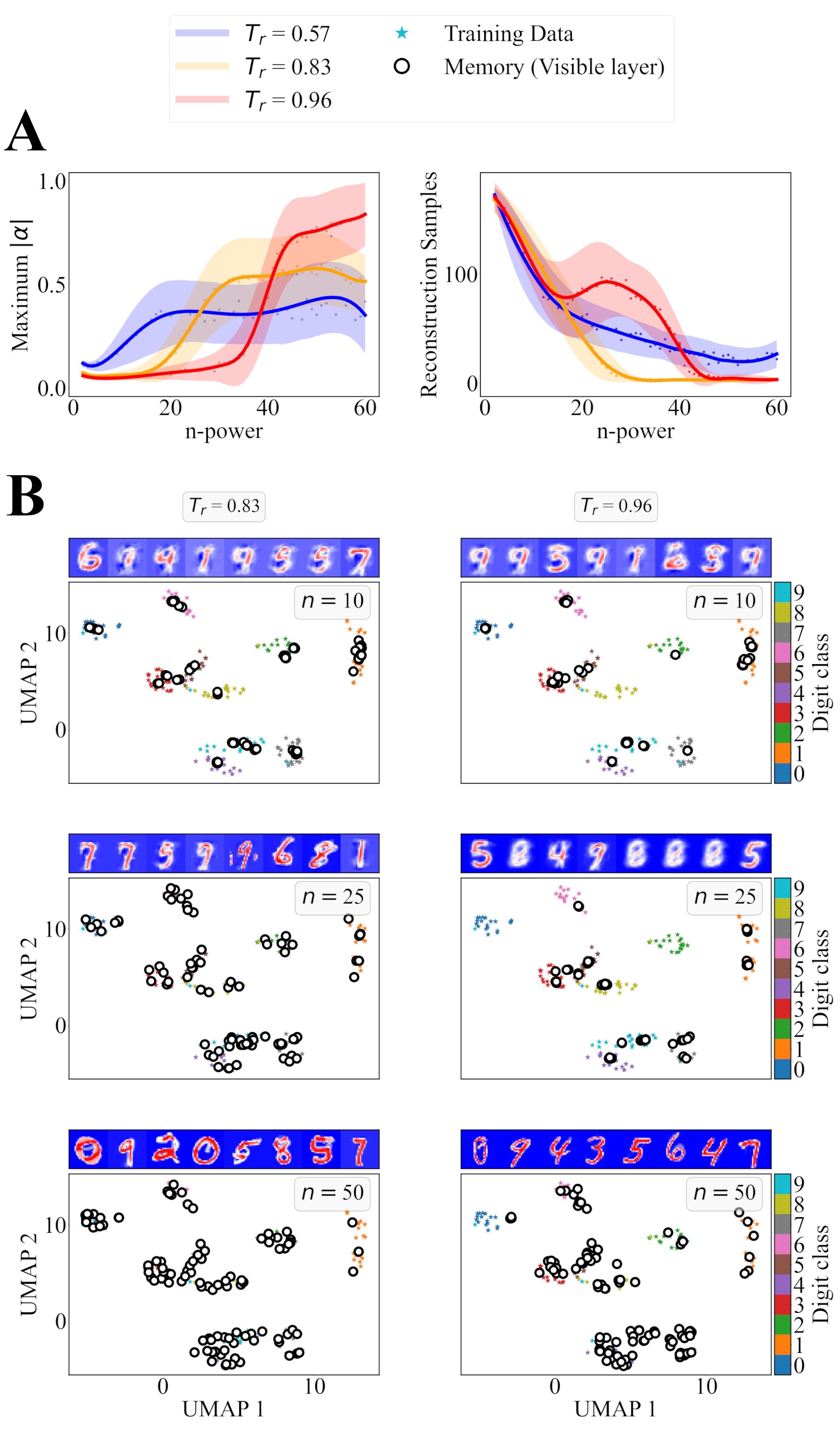}%{Figures (New, Merged)/Figure-Entropy.png}
  \caption{(A) Average of maximum $\alpha$ and average number of prototypes needed to reconstruct each memory, as a function of $n$ at the end of training, for different temperatures (B) Representation of the learnt memories for various $n,T_r$ using a common UMAP 2D embedding, computed with the entire MNIST training database. The training dataset is represented with points, colored according to their class, see the colorbar on the right. The memories at the end of training are represented by white disks, all learnt memories (100) are shown. A small sample of memories obtained is shown above each plot. For higher $n$, the memories are more spread and completely recover each cluster, indicating good coverage of the training set used. Other simulation parameters are the same as Fig. \ref{fig:fig0}.
}\label{fig:entropy}
\end{figure}

To better characterize the transition, we evaluated a few more metrics related to the  $\alpha_{\ket{i}}$. Fig. \ref{fig:entropy} A, left, shows the maximum value of $\alpha_{\ket{i}}$  at the end of the training, averaged over all memories in the system. We see at least two regimes. For smaller $n$, the maximum $\alpha$ is very small, indicating that no sample dominates, and thus a very distributed, or high entropy, encoding. As $n$ increases, there is a threshold in $n$, \textit{with possibly a discontinuous derivative}, from which the maximum $\alpha$ increases to significantly high values, thus defining prototypes.  In Fig. \ref{fig:entropy} A, right panel, we further show the average number of training samples necessary to reconstruct memories up to a small tolerance (see details in Supplement). Consistent with the behaviours on $\alpha$s, we see an approximately linear decrease for rescaled temperatures $0.57, 0.83$, up to a very low plateau, where few samples define the memory. Surprisingly, when the rescaled temperature gets closer to $1$, we observe an intermediate plateau for intermediate $n$, where on average about $50$ to $100$ samples are necessary to reconstruct a given memory.   To better visualize what happens here, and inspired by computational biology (e.g. the analysis of single cell RNA seq data \cite{Freedman2021,Yampolskaya2023}), we show memory samples as well as  memory locations using a UMAP embedding \cite{McInnes2018,Johnson2022} of the  MNIST dataset , Fig. \ref{fig:entropy} B (see details in the Supplement). While for high $n$ the memories are well spread in each cluster of the UMAP embedding, for smaller $n$ and higher $T_r$ the memories are typically more concentrated in the UMAP embedding, close to the center of each cluster, indicating that there typically is less variety in the memories.

\subsection*{Waddington learning dynamics in the prototype regime}

The overall simplicity of the results, especially in the prototype regime, motivates a study of the learning dynamics of the  internal memories  to see when and how memories converge.  We first follow and interpret the learning dynamics by visualizing the memories as a function of training epochs.  We visualize learning using the UMAP embedding of the  MNIST dataset  defined in the previous section (see Fig. \ref{fig:tree-learning} A for $n=3$ --feature mode--  and Fig. \ref{fig:tree-learning} D $n=30$ --prototype mode--, $T_r=0.85$, see also Supplementary Movies, and Supplementary Figures for other values of $n$). We show samples of memories at different epochs, ordered from $0$ to $9$ based on their dominating labels $l^\mu_d$ at the end of the training Fig. \ref{fig:tree-learning} B,E. We also show the number of digits correctly recognized as a function of the epochs in Fig. \ref{fig:tree-learning} C,F.

The learning dynamics in simulations for $n=3$ appear rather uniform, in the sense that the memories appear to distribute themselves  and 'diffuse' simultaneously across most digits as the number of epochs increases, Fig. \ref{fig:tree-learning} A-C (notice however that the UMAP embedding is difficult to interpret for small $n$ since memories do not necessarily correspond to well-defined digits).After an initial period where only digits of category $1$ are correctly recognized, around epoch 370 sub-sets of almost every  digit are correctly classified, and the number of properly classified digits in each category simply increases from there, Fig. \ref{fig:tree-learning} C. 

% A one-to-many dynamics. 

By contrast, the training dynamics for $n=30$ favours some specific digits at different epochs of learning, with clear sequential steps Fig. \ref{fig:tree-learning} D-F. Initially, all internal memories are almost identical and quickly converge towards a memory that looks like a $1$. Then, new digits (and corresponding memories) are sequentially learned, giving rise to trees in the UMAP embedding. For instance, in the simulation shown in  Fig. \ref{fig:tree-learning} E,  the digits $4$ are initially learnt around epoch $1000$, then $9,7$ and $5,6$ around epoch $1300$, and other digits later. The order of appearance of memories is consistent with the order in which digits are properly recognized, Fig. \ref{fig:tree-learning} F. While there is some variability depending on initial conditions and hyper parameters (compare e.g.  Fig  \ref{fig:tree-learning} and Supplementary Figures), the sequence of learning is reproducible from one simulation to another as $n$ is increased, and pairs of digits looking like one another are learned together. Typically, $1,4,9$ are learnt quickly first followed by $7$; then $6,5$; $8,3$ and lastly $0$ and $2$.

Looking in more detail at individual memories, the learning process at high $n$ is reminiscent of Waddington's proposal of landscape dynamics, consistent with the 'tree-like' structured observed in the UMAP embedding Fig. \ref{fig:tree-learning} D. Focusing on the first new digit appearing, we see that all initial $1$s start changing, identically, eventually looking like a mixture of digits, typically $1$ and $9$ (Fig. \ref{fig:tree-learning}E, epoch $900$) and embedded as an $8$ in the UMAP. This mixed state appears to be saddle in digit space that is approached through the self-organized dynamics of learning itself. Subsequently, a symmetry breaking  (or 'split') event occurs: while one subset of memories goes back to look like $1$, the other subset of memories first initially looks like some mixture of $4,7,9$  (epoch $1250$), which are saddles for subsequent splits. New digits are learnt when a subpopulation of  memories starts to resemble  mixtures with yet-to-learn digits, before splitting to acquire new digit identity : for instance the memories  in Fig \ref{fig:tree-learning} E eventually differentiating into $3$s and $5$s are identical up to epoch $1800$. Learning trajectories of memories are also canalized, in the sense that multiple internal memories converge together towards the same saddle -- the salient features of the learning dynamics, saddles, are robust. For instance, in the random samples of memories displayed in Fig \ref{fig:tree-learning} E,memories with the same eventual identity largely follow the same dynamical path in memory spaces, resembling each other at all epochs. It is only towards the very end of learning, when their final identity is fixed, that they specialize into different prototypes. This also confirms that, despite the high dimensionality of the system, the crucial splits in the learning dynamics are  low dimensional, close to canalized 'saddle' memories, and thus might be amenable to theoretical modeling.

\subsection*{Three-category systems recapitulate the phenomenology of the dynamics}

\begin{figure*}
  \centering
  \includegraphics[width=17.8cm]{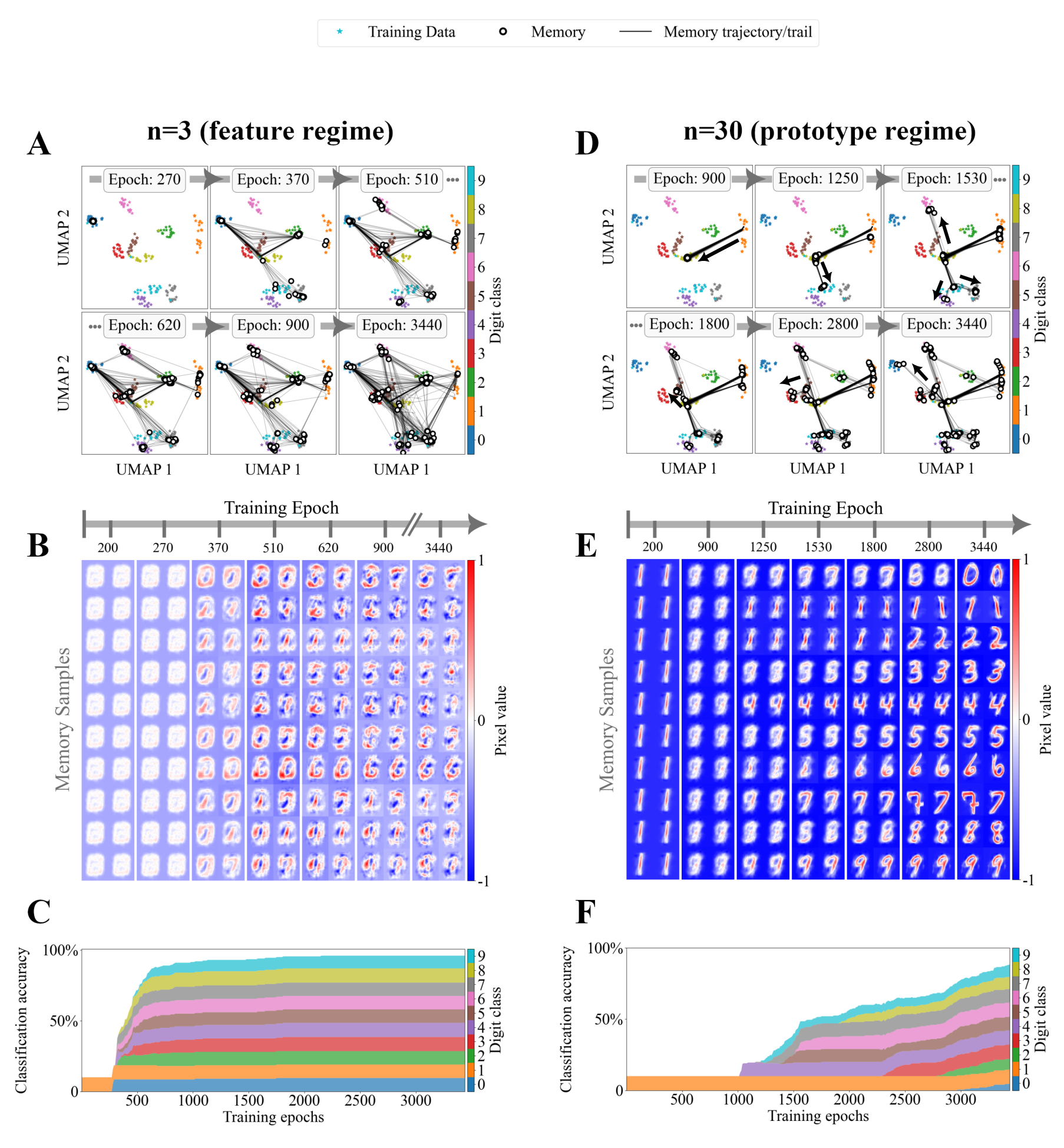}%{Figures (New, Merged)/Figure-1_merged.png}
  \caption{ Illustration of the learning dynamics, for two different values of $n$  Panels (A-C) illustrate what happens for $n=3$ (feature mode), while (D-F) are for $n=30$ (prototype mode).
  (A) The memories are transformed onto the UMAP latent space and plotted according to the epoch of training (indicated at top right corner of each subplot). A trajectory/trail of the learning path taken by the memories is also plotted in this space. The trajectories start at epoch 200 and end at the ‘current epoch’ of each subplot. B), a subset of the memories through training are shown. For the simulation behind this figure,  $T_r= 0.85$, other parameters similar to Fig. \ref{fig:fig0}. (C) Each coloured band height is proportional to the number of digits properly recognized as a function of the epochs, with color code indicated on the right    (D-F) Same as (A-C) for $n=30$ (prototype mode)
}\label{fig:tree-learning}
\end{figure*}

Motivated by the low dimensional aspects of digit learning at high $n$, we then proceeded to more deeply study simpler versions of the system, with the hope to capture the most interesting and generic features of the overall dynamics.

While the UMAP representation introduced in the previous section is useful for a qualitative understanding of the process, there are well-known problems with this method for quantitative analysis\cite{Patcher2023}. The $\alpha_{\ket{i}}$s provide a natural coordinate system, but they are difficult to visualize when there are too many samples. However, it turns out that coarse-graining multiple $\alpha_{\ket{i}}$ corresponding to the same digit allows for convenient representations of the dynamics close to saddles. For each memory $\ket{M^\mu}$ we thus define aggregated $\bar \alpha^\mu_d$s associated to a digit/category $d\in [0,9]$ such that :

\begin{equation}
\bar  \alpha^\mu_d=\sum_{\ket{i} \in \mathcal{T}_d} \alpha^\mu_{\ket{i}}
\end{equation}
where $\mathcal{T}_d$ defines all samples in the training sets labeled with digit/category $d$.

\begin{figure*}
  \centering
  %[width=11.4cm]
  \includegraphics[width=16.5cm]{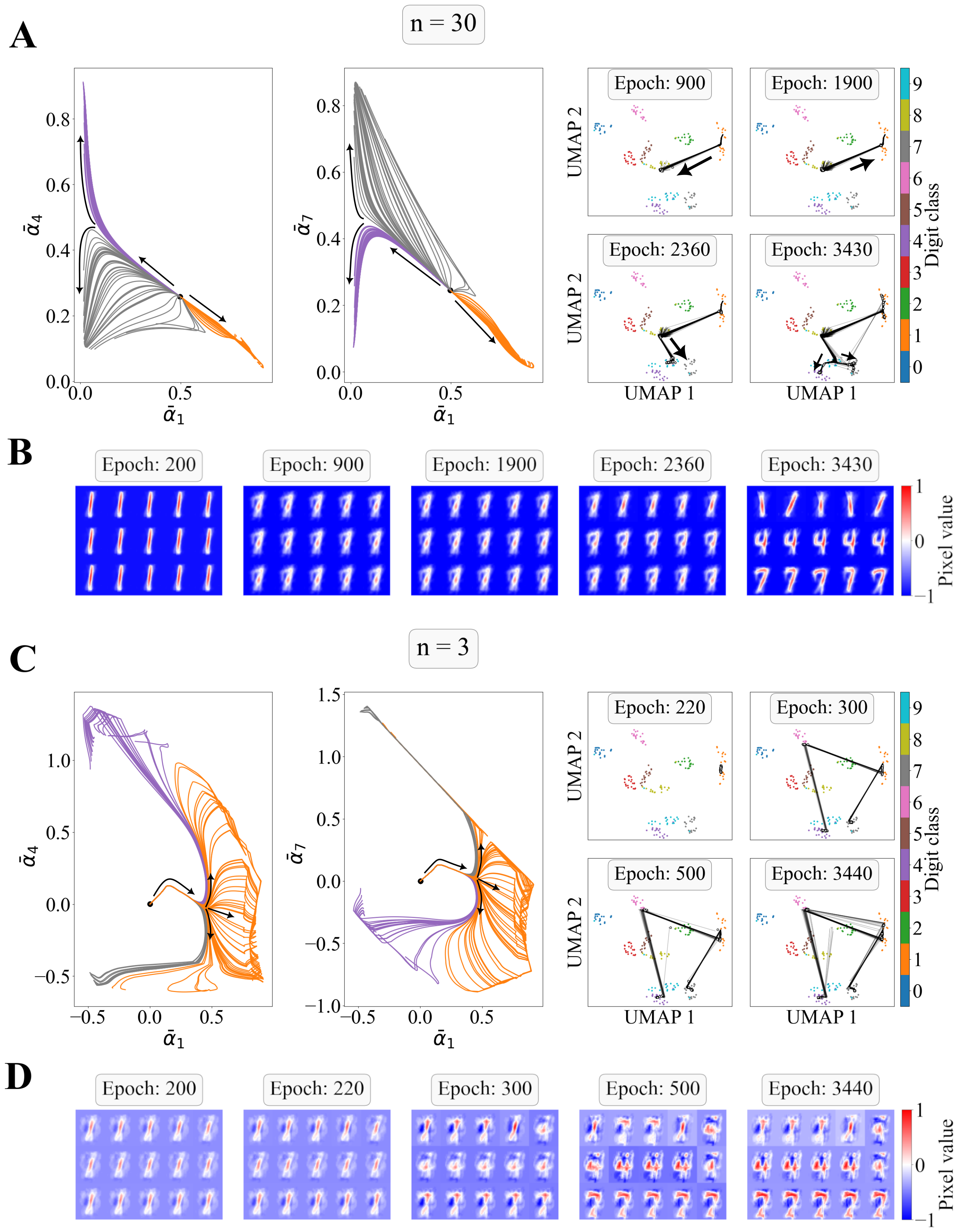} %{Figures (New)/Figure-2.png} % Made figure slightly smaller
  \caption{Splitting dynamics for high and low $n$. Here the system is trained on only three categories, $1,4,7$. We visualize trajectories in the similar UMAP embedding as in Fig. \ref{fig:tree-learning}  and display samples of the memories on the bottom row. We also display on the left the memory trajectory projected on the $\alpha$ coordinate, computed with the Moore-Penrose pseudo inverse. Trajectories are colored according to the dominating $\alpha$ at the end of the learning A) $n=30$. Two successive splits are clearly visible separating $1$ from $7,4$ (around epoch $2360$ as seen on the UMAP embedding), then $7$ from $4$ . B) $n=3$. The splits happen all around the same epoch $300$. Notice how there are many more connections between final memories in the UMAP embedding, and how the final memories look like mixtures of digits. }\label{fig:split}
\end{figure*}

The fundamental feature of sequential splitting can then be reproduced and visualized for simulations trained with only 3 categories of digits. In Fig. \ref{fig:split}, we represent learning trajectories of all memories for simulations where the training samples only contain $1,4,7$, using the same embedding as in Fig. \ref{fig:tree-learning}, for $n=3,30$ and $T_r=0.85$. In the UMAP space,  Fig. \ref{fig:split} A right, the steps in learning are virtually identical to Fig. \ref{fig:tree-learning}, with a first split localized in a $8$ cluster, and a second split in a $9$ cluster.  Such dynamics are all the more remarkable since neither $8$ nor $9$ sample digits are included in this reduced training set.  Convergence towards those clusters comes from the fact that the saddle prior to splitting indeed resembles an $8$, then $9$ . We also more clearly see in this simplified example how similar digits are initially identical, then eventually specialize into different prototypes, expanding within one cluster (late epochs in Fig \ref{fig:split} A).

In the $\bar \alpha_i$ space, Fig. \ref{fig:split} A left, one clearly sees a first convergence towards a saddle, then a split around $\bar \alpha_1 = 0.5$, sequentially followed by a second split around $\bar \alpha_7=\bar \alpha_4=0.5$, indicating that saddles are indeed  mixtures of digits. In between these two splits, the trajectories are largely canalized along the plane $\bar \alpha_1+\bar  2\alpha_4 \sim \bar \alpha_1+\bar  2\alpha_7 \sim 1$. We contrast this behaviour with what happens for low $n$ in the feature regime, Fig. \ref{fig:split} B, where the trajectories in the UMAP space do not correspond to clear steps and are not easily interpretable. In the  $\bar \alpha_i$ coordinates,  Fig. \ref{fig:split} B left, there is only one global split in a location where all $\bar \alpha_7$ and $\bar \alpha_4$s are small and the trajectories after the splits are spread out.  Such behaviors are generic and do not depend on the digits used: for instance in Supplementary Figures we further illustrate what happens for a different set of digits, $1,7,9$, with very similar properties.

\subsection*{Properties of the two-memory system}

The splitting of memories along canalized directions at saddles thus appears to fundamentally drive the dynamics of learning in the large $n$ regime. We hypothesized and checked that such splitting could be observed for systems with only two memories and two digit categories. Given that the memories all behave identically before the split and in each branch after the split, and that learning, globally, occurs through sequences of splits, we focus on the study of one single split in this section. To do so, we further reduced the system to two memories and only two input digits to discriminate, which allows for an analytical description of the convergence towards the saddles and of the subsequent split.

Before the split, calling $\ket{A},\ket{B}$ the two vectors corresponding to the two digits to discriminate, the (identical) memories can always be written as a linear combination of the two digits, defining

\begin{equation}
\ket{M}=\alpha_{\ket{A}} \ket{A}+\alpha_{\ket{B}}\ket{B}
\end{equation}

for the (common) memory before splitting. In Supplementary text and Supplementary Figures, we show that the labels are getting quickly correlated so that all labels get to $-1$ except the two labels corresponding to digits $A,B$ such that $l_A=-l_B$, subsequently called $\ell$. The learning dynamics act on the $\alpha$s and $\ell$ .

Memories are normalized at each learning epoch, and in particular one can show that due to normalization the system quickly localizes in a region where
\begin{equation}
|\alpha_{\ket{A}}|+ |\alpha_{\ket{B}}|=1
\end{equation}
where the actual sign of each term in this equation depends on the sign of $\ell$ (see Supplementary calculations, and Supplementary Figures). So, in the end, the dynamics towards the saddle are bi-dimensional, e.g. we  can eliminate $\alpha_{\ket{B}}$ to express the entire dynamics as a function of $\ell$ and $\alpha_{\ket{A}}=\alpha$.

We then carried out an effective nullcline analysis. We can write two effective equations (accounting for the effect of the normalization for the dynamics) for $\alpha$ and $\ell$. Putting them to zero defines two nullclines, respectively called 'memory' and 'label' nullclines, see Fig. \ref{fig:bifurcation} A and Supplement. The full dynamics of learning towards the saddle qualitatively follow the 2D vector fields defined by those nullclines, Fig. \ref{fig:bifurcation} B-D. More details and comparisons with the full dynamics are shown in the Supplementary Figures, we also show several examples of learning in the 2D plane in Supplementary Movies. Importantly, the fixed points are the invariant vectors for the normalization, i.e. such that $ \ket{\Delta M} = \Delta \alpha_{\ket{A}} \ket{A}+ \Delta \alpha_{\ket{B}} \ket{B}$ is proportional to the initial memory $\ket{M}$.

\subsection*{Bifurcation diagram define three different regimes for saddles}

\begin{figure*}
  \centering
  % [width=17.8cm]
  \includegraphics[width=14.8cm]{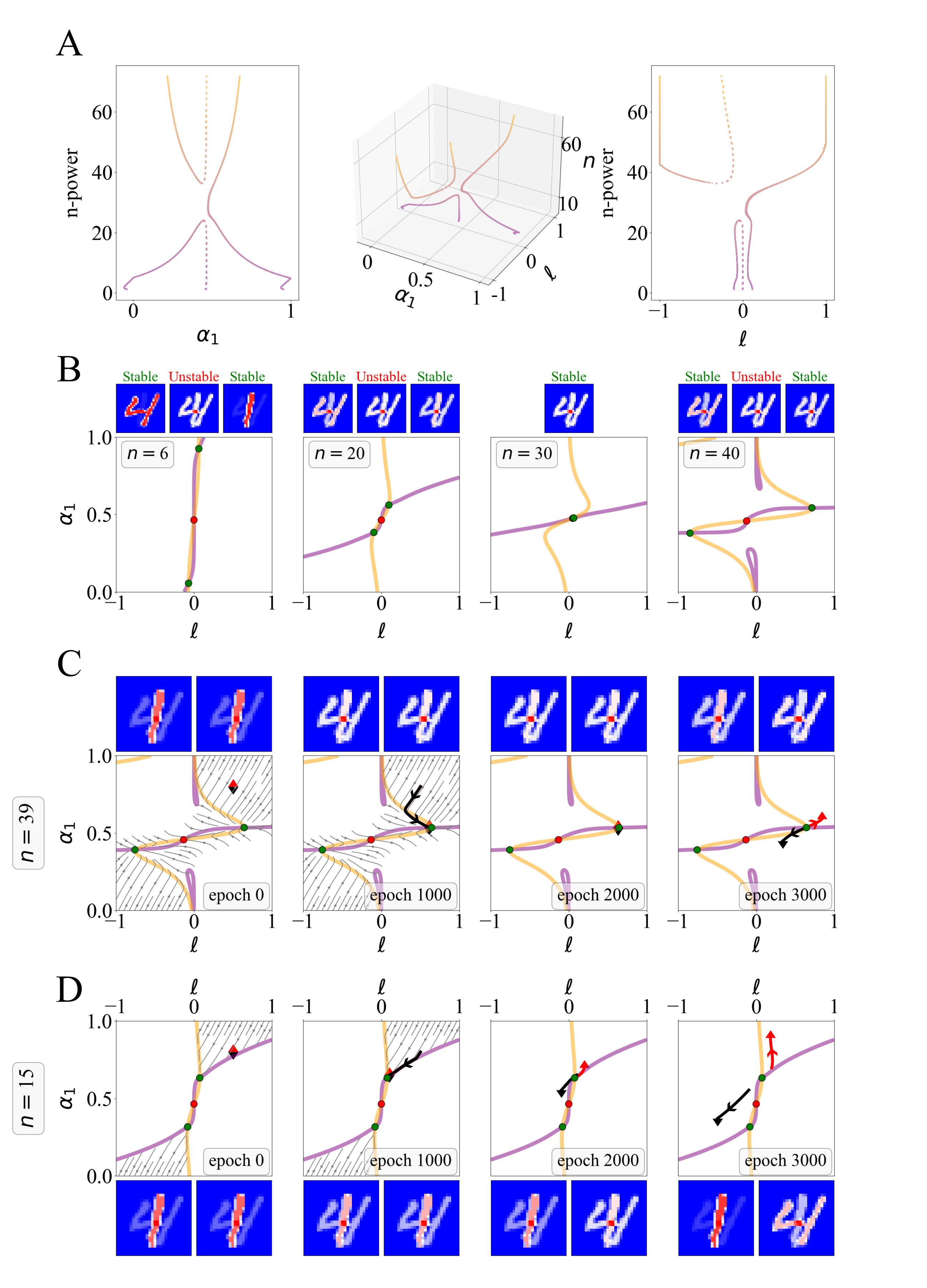}%{Figures (New, Merged)/Figure-3_merged.png}
  \caption{Bifurcation  and dynamics for the two-memory system.  In A), we show the bifurcation diagram of a 2-memory system with 2 identical memories, using $n$ as a control parameter, and $\alpha$ and $\ell$ to show the fixed points. The center plot of A) is the complete 3D bifurcation diagram of this system, the left and right plots show projections on planes of constant $\alpha$ and $\ell$ respectively. In B), the fixed points as a function of $n$ are illustrated on the top and the nullclines (memory: purple, label: orange) are plotted below. Note that in the presence of 3 fixed points, the ‘center’ point is unstable. (C) Splitting dynamics for the two-memory system. One displays the nullcline corresponding to the saddle, flow lines corresponding to the analytically reduced dynamics, and we plot the actual dynamics in the $\alpha,\ell$ plane for  $n=39$ D) same with $n=15$. All simulations in this figure are for a 2-memory system with $T_r=0.89$, for training data with two sample digits $1,4$ corresponding to $\ket{A}, \ket{B}$ with $\braket{A|A}=753, \braket{A|B}=494$ and $\braket{B|B}=719$.
}\label{fig:bifurcation}
\end{figure*}

The two nullclines are sigmoidal, and one observes three qualitatively different regimes going from bistability to monostability to again bistability as $n$ is increased, see e.g. Fig. \ref{fig:bifurcation} A using digits $1,4$ as examples. Multistability in this context means that, depending on the initial conditions, the system first splits at a different saddle. The fixed points themselves look like digits for low $n$ (first bistable region), then like a mixture of digits in the monostable phase for intermediate $n$, and again gradually look like single digits in the second bistable region as $n$ is further increased.

Fig. \ref{fig:bifurcation} B illustrates nullclines for various values of $n$.
For very low $n$, both nullclines are almost vertical, parallel to the $\alpha$ axis, defining a critical valley at a fixed value of $\ell$, but where future splits between memories are very sensitive to noise and initial conditions. This is consistent with the stochasticity observed in the full system (see Supplementary Movies). Intuitively, vertical nullclines (close to $0$) mean that the system is unable to label the internal memories properly: all labels stay close to $0$ (as can be seen on the bifurcation diagram,  Fig. \ref{fig:bifurcation} A right panel, low $n$)  and there are only weak biases in one direction or the other depending on the fixed point.

As $n$ is increased  Fig. \ref{fig:bifurcation} B, $n=20$, the slope of the memory nullcline decreases first, so that the system has three critical points, two stable and one unstable. Interestingly, because the label nullcline stays relatively vertical, they correspond to relatively high and low values for $\alpha$ parameters, meaning that the fixed  points look very much like one of the initial digits but with some background of the other one. Notice however that the $\ell$ stay low, meaning that while the memories are relatively well defined, their labels are not.

As $n$ is further increased, Fig. \ref{fig:bifurcation} B, $n=30$, those stable fixed points disappear through a saddle-node bifurcation, and remarkably, only the fixed point close to $\alpha=0.5, \ell=0$ survives.  Intuitively,  the system gets more selective in its recognition of proper digits, so that any mixture of digits can (and should) not be categorized as either category. The perfect mixture gets an ambiguous $\ell=0$ label, which is the only fixed point surviving. In Supplementary Figures, we show that the system is very close to a pitchfork bifurcation, due to the fact that most pixels in the initial pictures take values close to $1$ or $-1$, so that $\braket{A|A} \sim \braket{B|B}$.

As $n$ is further increased, a new saddle-node bifurcation occurs, where $\ell$ values initially become significantly higher than $0$. This comes from the fact that the memory nullclines become increasingly horizontal close to $0$.  This is the more intuitive regime in the generalized Hopfield energy landscape: as $n$ is increased, the 'energy' of the saddle point gets very frustrated between the two digits, so that the only critical point for the one memory system should be exactly a superposition of both digits corresponding to $\alpha \sim 0.5$. However, there is a slight bias in the loss function due to the label, so one gets two stable fixed points, in which one digit slightly dominates the other in the memory, but with an unambiguous categorical label $\ell$. As $n$ is further increased, those biases on the labels allow for further symmetry breaking between the two fixed points, which eventually look increasingly like 'pure' digits.

Examples of the dynamics of two memories are shown in Fig. \ref{fig:bifurcation} C-D in the $\alpha,\ell$ coordinate. The dynamics of the full system are initially structured by the $\alpha,\ell$ nullclines, and both memories indeed converge to the (closest) saddle, before splitting. After splitting, one needs more dimensions to describe the full dynamics, but some small approximation (defining small deviation from the saddle $\delta \alpha, \delta \ell$) can be done to compute and visualize what happens close to the saddle. In Supplementary Figures, we compare the simulated trajectories with our analytical computations of the small deviations close to the saddle, with excellent agreement.

% The image has margins (left, right)

\subsection*{Feature-to-prototype transition for the two-memory system}

\begin{figure}
  \centering
  \includegraphics[width=\linewidth]{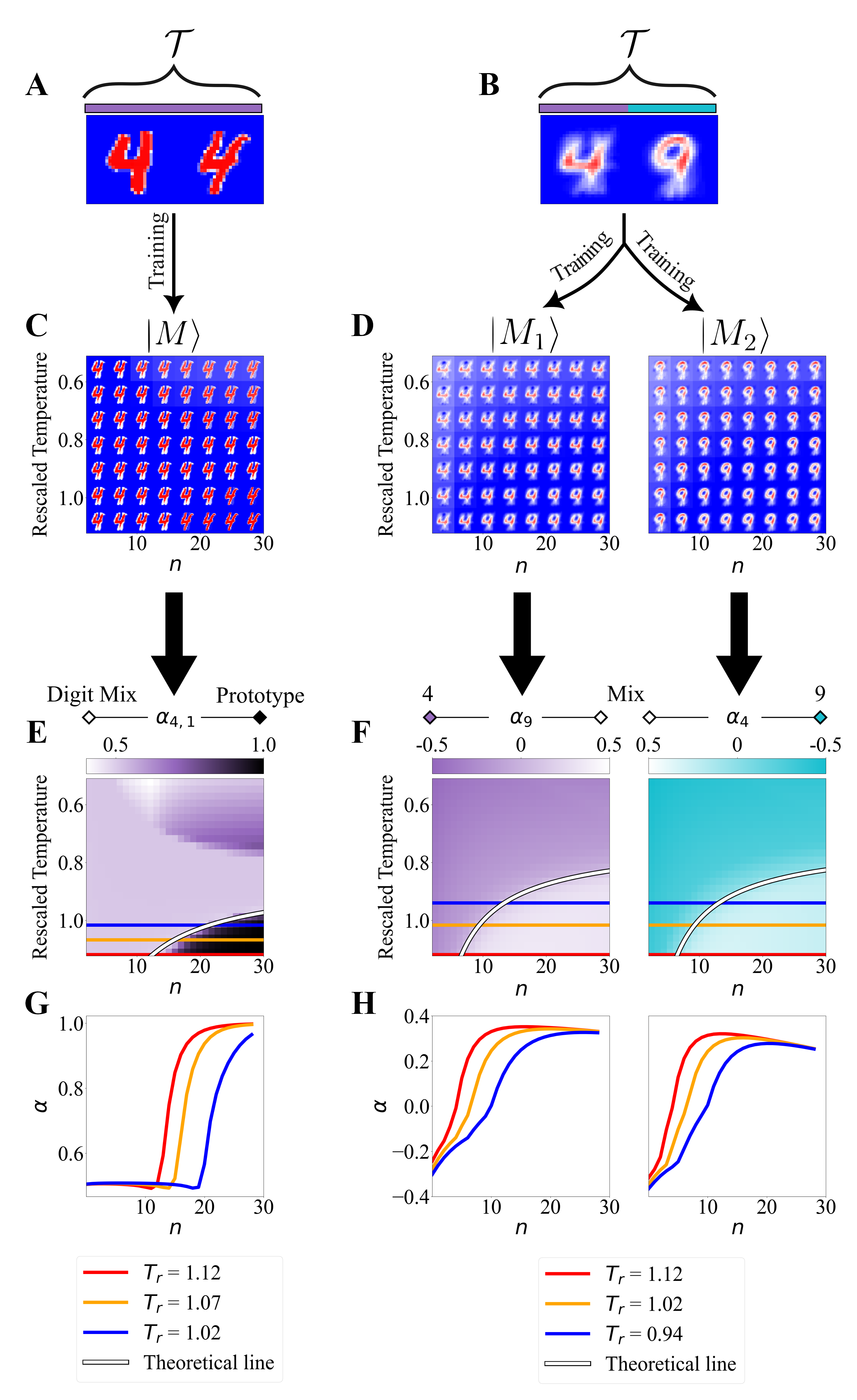}
  \caption{Two memory system recapitulates a feature-to-prototype transition (A) Training set used to study the one-memory intra-digit system (B) the training set for the two-memory inter-digit system (C-E-G) Intra-digit specification. We show the final state of the memory as a function of $n,T_r$ (C), the corresponding $\alpha_4$ (E), and curves of $\alpha_4$ as a function of $n$ for 3 high temperatures (G). The white line on (E) indicates when the score of a perfectly mixed digit is the same as the score of either digit in (A), coinciding with the feature-to-prototype transition. (D-F-H) Inter-digit specification. We show the final state of memory 1 (left) and 2 (right) as a function of $n,T_r$ (D), the corresponding $\alpha_9$ of memory 1 (left) and $\alpha_4$ of memory 2 (right) (F), and curves of $\alpha_9$ (left) and $\alpha_4$ (right) as a function of $n$ for 3 high temperatures (H). The white line on (F) indicates when the score of a perfectly mixed digit is the same as the score of either digit in (B), coinciding with the feature-to-prototype transition. }\label{fig:twomemory}
\end{figure}

While the bifurcation diagram described above depends on $n$, it relates to saddles in the learning dynamics, but does not  clearly inform on the feature-to-prototype transitions which relates to the fixed points. Such fixed points can however be computed analytically for the two-digit/two-memory system. Figs. \ref{fig:twomemory} illustrates final memories for different values of $n,T_r$ for two cases : one where digits are of the same category (so having the same $\ell$), Fig. \ref{fig:twomemory} A, C, E, and the other where digits are of different categories (so opposite $\ell$) , Fig. \ref{fig:twomemory} B, D, E . We illustrate how the $\alpha$s of the fixed points are changing as a function of $n,T_r$ for both cases. Clearly, as both parameters are increased, the final states change from a mixture of both digits (looking like features), to a more defined digit (looking like a prototype). Consistently, one gets from fixed points with one positive and one negative $\alpha$, to fixed points  where only one $\alpha$ dominates (from top left to bottom right in Figs. \ref{fig:twomemory} C, E and \ref{fig:twomemory} D, F).

It is also possible to analytically compute the transition line where a single digit yields the same cost as a perfect mixture of digits, presumably corresponding to the feature-to-prototype transition (see Supplement). We compare it to actual simulations, with excellent agreement (white line in Figs. \ref{fig:twomemory} E and \ref{fig:twomemory} F)

For the intra-digit discrimination (i.e. same category), one gets pure prototypes for high $n,T_r$, in the sense that the $\alpha$ contribution from one of the digits completely vanishes. Also, there is a very clear threshold in $n$ from which one $\alpha$ suddenly increases above $0.5$, thus defining two regimes, and reminiscent of the behaviour of the maximum of $\alpha$ displayed in Fig. \ref{fig:entropy} A. However, for the inter-digit discrimination (i.e. different categories), the behaviour of both $\alpha$s is smooth and we do not see a clear separation between two regimes. Furthermore, even for big $n,T_r$, one still sees a small but non-zero contribution of both digits in the fixed point. So one never gets to a 'pure' digit representation as expected from a true prototype: rather the fixed points look like some mixed saddles in the high and low $n$ limit in Fig. \ref{fig:split}.

We thus conclude that two-memory systems recapitulate some of the salient aspects of the feature-to-prototype transition observed in the full system, suggesting it is not an aspect of the learning that is due to the high number of memories in the system but that, surprisingly, the different 'phases' in $\alpha$ are more visible for intra-digit categories than for inter-digit categories. This is however consistent with the observation made on the full system that initially similar memories differentiate into prototypes late in the training : it is likely only when two memories have the same label that the intra-digit split is happening, putting almost all $\alpha$s  to $0$ and thus defining a proper prototype.

\subsection*{Learning and final memory statistics in the expanded system are structured by bifurcations}

We have thus identified two aspects of the two-memory system: a bifurcation diagram structuring the saddles visited during learning, and a feature-to-prototype transition on the final states of the two-memory system. Importantly, while those two aspects are correlated (because they both depend on $n,T_r$), they are independent in the sense that one does not need the knowledge of the bifurcation diagram of the saddles to compute the final states of the memories. We know that the feature-to-prototype transition generalizes from systems with few memories to bigger ones, but it is not clear if the number or nature of saddles matter in any way for bigger systems. We thus now re-expand the number of memories and samples to study how/if these properties exhibited for the two-memory system generalize.

\begin{figure}
  \centering

  \includegraphics[width=\linewidth]{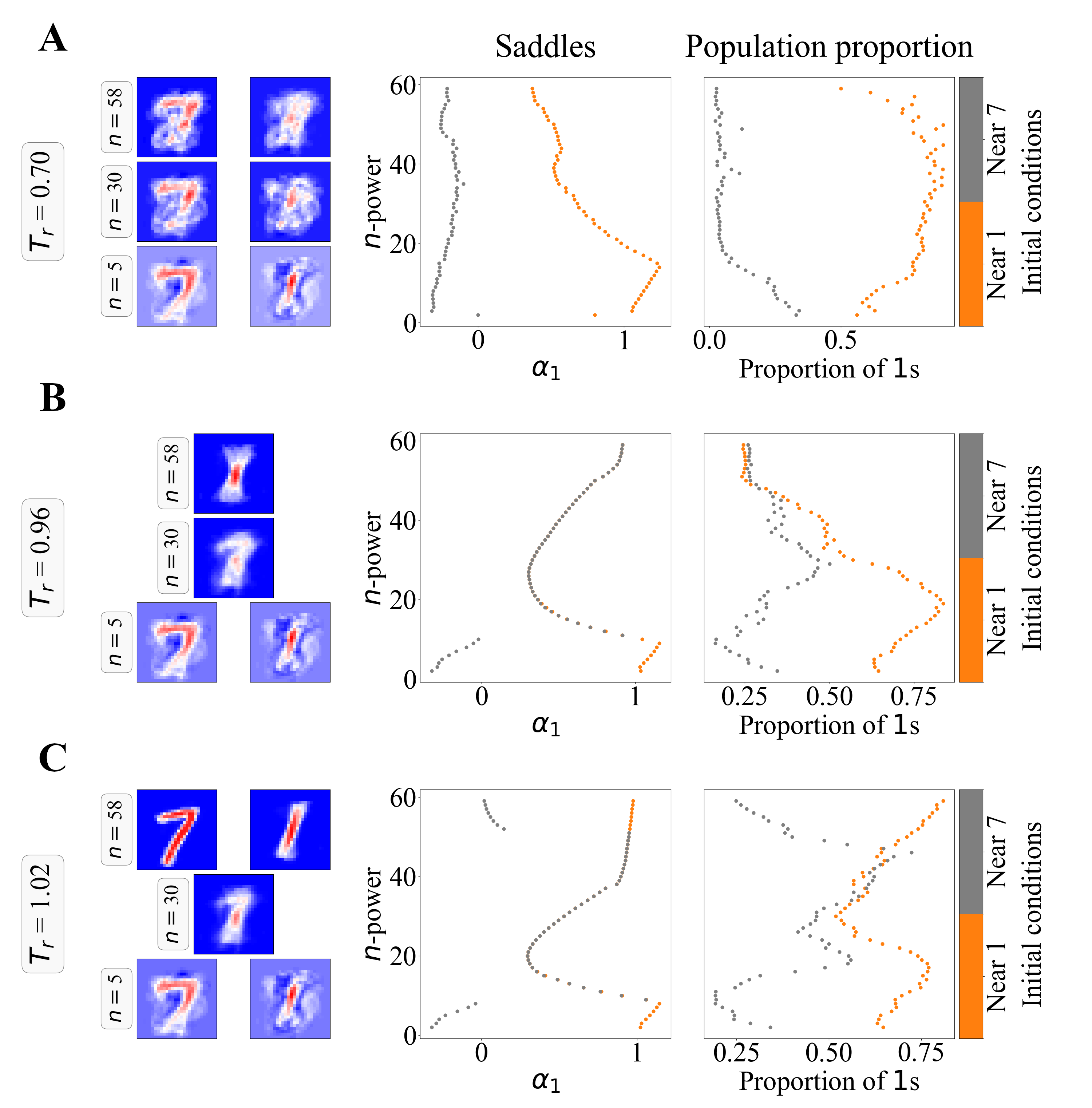} 
  \caption{Bifurcation of saddles correlates to changes of proportions. We show results of simulations respectively initialized close to a $1$ sample digit (orange dots) and  close to a $7$ sample digit (grey dots). From left to right, we show examples of saddles reached in the left panel, $\alpha_1$ coordinates of the saddles in the middle panel , and the relative proportion of $1$ vs $7$ in  the right panel as a function of $n$. (A) $T_r=0.7$ (B) $T_r=0.96$ (C) $T_r=1.02$ }\label{fig:general_bifurcation}
\end{figure}

Because of the combinatorial explosions of possible memories and saddles, we move back to numerical explorations to explicitly study the influence of saddles on learning. We first restrict ourselves to a system with 100 memories, 3 categories ($1,7,9$) and 20 samples per category. To explore different learning trajectories, possibly corresponding to different saddles, we initialized the memories close to different digits, then performed learning for different temperatures Fig.  \ref{fig:general_bifurcation} A-C and focus on the $1$ vs $7$ discrimination. We first recovered multiple saddles in learning: we observe again at least three regimes for high temperature $T_r \sim 1$, with at least two bistability regions for low and high $n$, and an intermediate region where the system converges towards a single saddle.

So a similar bifurcation diagram to the two-memory case is also observed for systems with many more memories. Importantly, we also observe that the \textit{proportion} of final memories of a given identity depends on initial conditions, and thus on the saddle first visited Fig. \ref{fig:general_bifurcation}A-C), a property that can not be observed on systems with two memories since each memory stabilizes to either fixed point. Consistent with this, the proportion of $1$ and $7$ in the final memories is correlated to the respective values of $\alpha_1$ and $\alpha_7$ at the saddle visited. For instance, for $T_r \sim 1$, Fig. \ref{fig:general_bifurcation}C, because there are more saddles for small and big $n$, when the system is initialized close to $1$ (resp $7$), the proportion of $1$  (resp $7$) in the final memories is very high for most parameters, while it is very low if the system is initialized close to $7$ (resp $1$). Only for intermediate $n$, when there is only one saddle visited irrespective of the initial conditions, do we observe a balance between $1$ and $7$s in the final memories, when the (unique) saddle is a mixture of $1$ and $7$.  All in all, the final state reached after learning thus appears to be indeed path-dependent, and the number and nature of saddles are relevant for the final statistics of the memories learned.

\begin{figure}
  \centering

  \includegraphics[width=\linewidth]{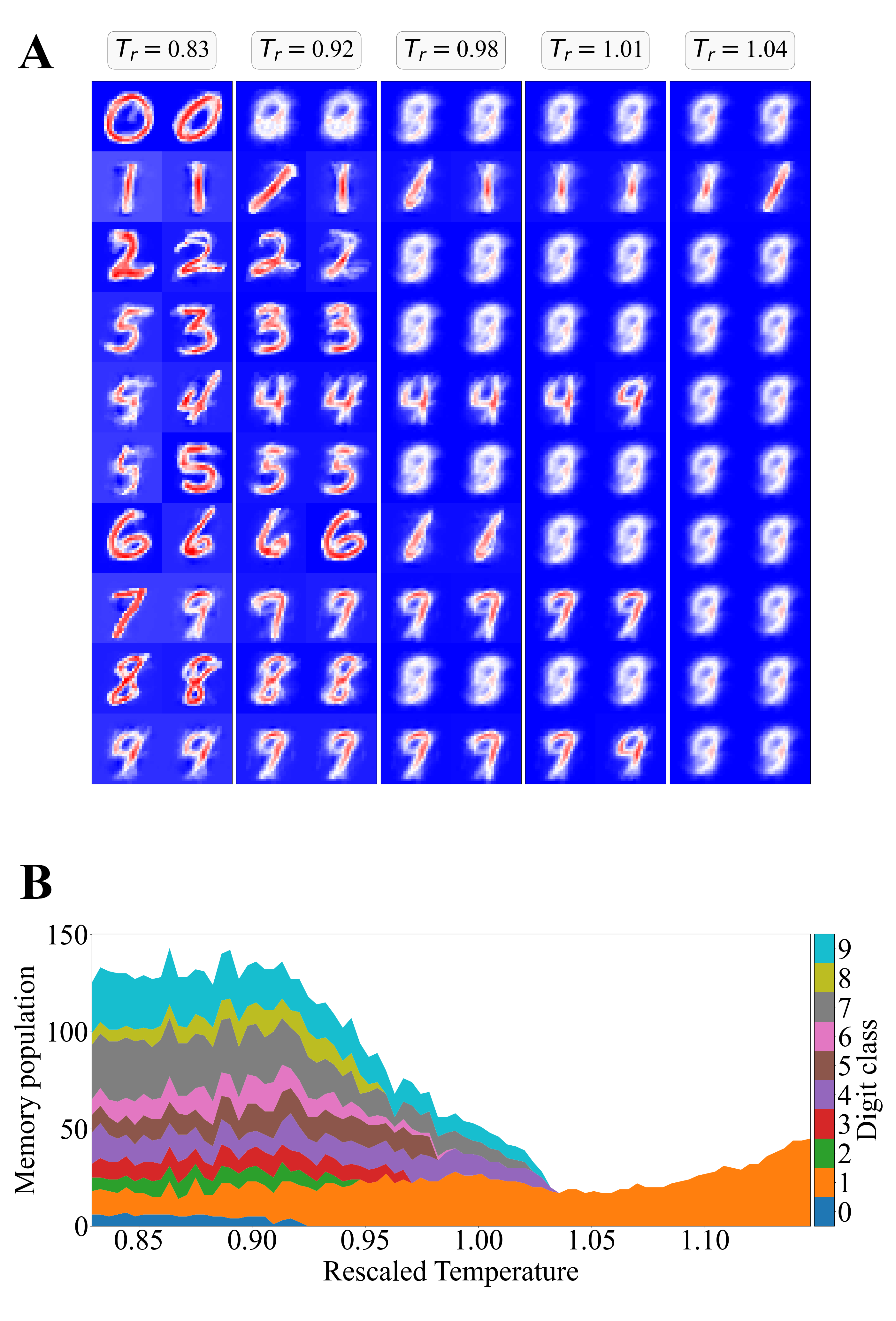} 
  \caption{ Training for varying temperature for $n=25$. As the temperature increases, some categories are no longer properly recognized. (A) Samples of final memories for different temperatures. (B) Population for each memory (as inferred from their label) as a function of Temperature after training. }\label{fig:temperature}
\end{figure}

Importantly, we also see a clear $T_r$ dependency on both the bifurcation diagram and the proportion of memories, Fig. \ref{fig:general_bifurcation}A-C. We thus examined what happens for the full system as we vary $T_r$, in a range of $n$ where we expect few saddles (typically $n\in [20,30]$ according to Fig. \ref{fig:general_bifurcation}B-C). In Fig. \ref{fig:temperature}, we show samples and the number of memories for each category after training, as a function of temperature (using trained labels as a proxy, see Supplementary Movies. For lower $T_r \lesssim 0.85$, all digits are represented in the memories. However, as $T_r$ increases, some memories disappear and instead start looking like the typical first saddle observed during training  Fig. \ref{fig:temperature} A. Remarkably, as $T_r$ is increased, there is a clear order in the disappearance of memories : first $0$ and $2$, then $3,8$; $5,6$, and $4,7,9$, leaving only $1$ in the end when $T_r>1$. This order essentially is the inverted order of digits learnt during one instance of successful training of the system, compare e.g. Fig. \ref{fig:tree-learning} F. This suggests that as $T_r$ is increased, the system is no longer able to split in some directions. Instead, it remains stuck at saddles close to root of the tree displayed in Fig. \ref{fig:tree-learning} C as $T_r$ is increased.

\section*{Discussion}

Artificial neural networks offer examples of non-trivial self-organizing dynamics, and as such constitute easy-to-simulate comparison points to real-life complex networks.  In this work, we studied the learning dynamics of generalized Hopfield networks trained to classify MNIST, previously known to exist in two broadly defined regimes (feature and prototype based) as a function of hyper parameters characterizing the non-linearity of the energy landscape. 

We visualized and studied the learning dynamics of such networks. We especially focused on the prototype dynamics(for higher $n$) where learning is most reproducible:  memories localize first at 'saddles', corresponding to mixtures of digits, subsequently memories 'split', before reaching either a new saddle or specializing into an actual digit. Learning is canalized, in the sense that memories with the same eventual digit identity largely follow the same pathway in memory space during learning, and only specialize into different prototypes in later epochs. Consistent with this, the order in which memories are learned is reproducible across simulations for large $n$.  

We explore the dynamics of this complex learning scheme through simpler versions of the systems with fewer memories, establishing the local low-dimensionality of the process, and allowing us to better characterize the most salient properties of the learning dynamics. We reveal that the number and nature of possible saddles of the dynamics depends on $n,T_r$, in particular, we see for intermediate $n$ and higher temperature a regime where there are a smaller number of saddles in the dynamics of learning, which themselves appear and disappear via saddle-node bifurcations at both lower and higher $n$, respectively. 

Our results suggest the following view of the learning dynamics in the prototype regime: for lower $n,T_r$, there are multiple saddles, and thus as shown in Fig. \ref{fig:tree-learning},  we generically observe more path dependency and randomness in learning, leading to different proportions of memories with a given digit identity. However, for mid range $n$ (typically $n \in [20,30]$ for $T_r \sim 0.89$), the system is in a regime where fewer saddles  exist, due to the first saddle-node bifurcation (Fig. \ref{fig:bifurcation} B, $n\sim 30$). In this regime, owing to the paucity of saddles, learning is expected to be much more canalized (less dependent on initial conditions and noise). This is  consistent with the plateau observed in Fig. \ref{fig:entropy} A right for higher temperatures : where memories are 'trapped' at various saddles, as visible in the UMAP plot in Fig. \ref{fig:entropy} C  where memories do not distribute well within one digit cluster. It is then only for even bigger $n$ ($n>30$) that more saddles appear again. Importantly, these saddles correspond to mixtures of digits, but 'biased' towards one digit. This allows the learning to have more path dependency, but locally (i.e. within a digit category), giving rise to many more prototypes as final states. This provides a rationale as to why the memories in the UMAP plot in Fig. \ref{fig:entropy} B are more spread out for larger $n$. Manifestly, the location and nature of saddles influence the proportion of final memories in a given digit category. Interestingly, as we increase $T_r$, the system gets increasingly stuck in saddles, following the canalized pathway of learning. This is consistent with the observation that the learning dynamics of a digit category is low-dimensional and canalized along a small number of well-defined directions, with the detrimental effect that it is also 'easy' to sequentially block those directions during learning. This does suggest that there is a 'sweet spot' to ensure a desirable prototype distribution over all digits : a low number of saddles between digit categories is needed to have a good number of memories corresponding to all digits, but if the temperature is too high learning can get stuck at those saddles. 

We wish to highlight that the feature-to-prototype transition is not restricted to systems with a large number of memories and, importantly, the pure prototype regime where only a single digit ($\alpha$) dominates appears to come from intra-digit classification, rather than inter-digit classification.  This is consistent with the fact that memories with the same eventual label largely appear to change together, as can be seen in Fig. \ref{fig:tree-learning} E and Supplementary Movies. It is only when memories have the same final label that intra-digit interaction ensues and that they become pure prototypes.

How general are our results? Of note, generalized Hopfield networks have been related to both diffusion models \cite{Hoover2023} and transformers \cite{Ramsauer2020}, the architecture at the core of the success of current Large Language Models, such as chatGPT.  In particular, Ramsauer \textit{et al.}  \cite{Ramsauer2020} use an explicit energy with an exponential for the function $F$ defined in Eq. \ref{eq:dfn_F} , and thus do not have the equivalent of a hyper-parameter $n$.  They did not study the dynamics of learning, but studied the types of attractors, and observed both single patterns (corresponding to prototypes) and mixtures of 'close' patterns, which within our framework would likely correspond to saddles. However, they did not explore the anticipated bifurcations. It is unclear if there is any way those saddles would split into well-defined prototypes like in our case. Interestingly, the transformer learning rule in their context corresponds to changes in $\alpha$, in our context. They observed the distribution of activation patterns in layers of actual transformers (similar to distribution of $\alpha$s) and found changes reminiscent of what is observed here when $n$ is increased, e.g. with some mid-late layers activated by very few patterns, while early layers are broadly activated by many patterns. This suggests that a feature-to-prototype transition might be a generic property happening within different layers of neural networks, which provides further motivation for case-studies of more tractable systems as we have explored here.

The tension between the high-dimensions of the sample space and low dimension of the dynamics is very reminiscent of what happens in complex (biological) systems, a primary motivation for the present study. In the prototype regime, memories with the same terminal identity differentiate sequentially, moving from one saddle to the next, before splitting. This resembles the classical landscape exploration we anticipate, \textit{à la Waddington}, during the time course of a differentiating cell. Extending this, we anticipate that  dynamical systems theory can be used as an overarching framework to parametrize learning systems (such as the change of saddles, Fig. \ref{fig:bifurcation} ). In particular, we expect the current new wave of applications of 'catastrophe theory' \cite{Thom} applied to the description of biological systems to possibly extend to machine learning dynamics\cite{Rand2021, Francois2023}. Within this framework, it has been suggested that cellular differentiation is largely driven by external morphogens, acting as external control parameters 'tilting' the landscape or inducing bifurcations during differentiation \cite{Rand2021}. Here, we notice that, upon the increase of temperature $T_r$, the system stays stuck in the saddles normally visited during learning. This is reminiscent of a well-known, general experimental fact : in culture, intermediate progenitor (stem) cellular states can be maintained indefinitely upon constant exposure to the proper combination of morphogens. The parameter $T_r$ might thus play a role in modulating the tilts so as to maintain 'stemness'. Then, for low enough $T_r$, the differentiation we observe is completely \textit{self-organized}. We observe successive symmetry breaking leading to splits (Fig. \ref{fig:split}), so that, from a single memory standpoint, the landscape is not fixed, but rather changing due to the interaction with other memories. This is reminiscent of the second (but less well-known) picture proposed by Waddington, of weights and levers as epigenetic controls of the shape of the landscape.  It is thus striking that in our reduced two-memory system, dynamics are controlled by an interplay between memories ($\alpha$) and labels ($\ell$). This suggests that the labels might play a similar role to the weights underlying Waddington landscapes, here with more direct interpretability since labels directly correspond to memory identities. If memories are analogues to gene expression states, labels might  correspond to less labile variables phenotypically defining a cellular state (chromatin marks, morphologies), that in turn control morphogens.

Of note, Matsushita et al. \cite{Matsushita2022,Matsushita2023} have recently proposed a new model for cell differentiation based on random network theory (modelling gene interactions), combined with slowly changing epigenetic variables. Such systems generically display oscillations, before 'localizing' their dynamics towards multiple steady states. They suggest that 'stem cells' should not be considered like quasi-static states, but rather oscillatory states.  This offers an alternative view, again inspired by dynamical systems theory, and a comparison point to the sequential differentiation dynamics in the high $n$ limit described by us.  Interestingly, their separation between genes and epigenetic variables is again reminiscent of the separation between memories and labels that we study here. This further suggests that differentiation phenomena might be best described and understood with multi-layered dynamical systems, motivating further theoretical studies of this class of systems similar to what we propose here.

Differentiation dynamics thus apply beyond biology, and we propose that machine learning offers a new route to studying this phenomenon, both from a theoretical and experimental (numerical) standpoint. Going the other way, one can imagine that the feature-to-prototype transition, or the multiple regimes of learning that we observe, might also apply to differentiation biology. For instance, do more random routes to (de)differentiation exist, correspond to the low $n$ regime ? -- providing counter examples to Waddington dynamics in restricted contexts. The two-way bridge that we are proposing, between the dynamics of cellular differentiation and machine learning, legitimizes the use of (Generalized) Hopfield Networks \cite{Lang2014} as direct models of differentiation biology, with interpretable handles (e.g. we mentioned the role of $T_r$ to maintain the systems close to saddles).
Differentiation or evolution of attractors, similar to the 'saddle-node bifurcations of saddles' we observe  as $n$ is varied (see e.g. bifurcation diagram in Fig. \ref{fig:bifurcation}), might further offer natural roads for biological/neuronal learning if biological hyper parameters equivalent to $n,T_r$ are dynamically varied. 

\section*{Appendix}

We study the classification problem : a cost function is defined 
\begin{equation}
C=\sum_{\ket{\sigma} \in \mathrm{Training}} \sum_d (c_d (\ket{\sigma})-t_d(\ket{\sigma}))^{2m}
\end{equation}
where $\ket \sigma$ is the (vectorial) input presented, \(c_d (\ket{\sigma})\) a categorical score of the input associated with category $d$  computed by the network, and  $t_d(\ket{\sigma}$ the actual  (or target) score of the vectorial input $\ket \sigma$ . The goal of learning is to minimize the cost function, so that after learning each input is correctly categorized. Very concretely, we focus on the standard digit classification using MNIST \cite{lecun1998mnist}.  $\ket \sigma$ then represents a picture  of size $28 \times 28$ pixels, so dimension $784=28 \times 28$  . $d$ represents a digit category ($0$ to $9$), $t_d(\ket{\sigma})=1$ if $\ket \sigma$ belongs to category $d$ and $-1$ otherwise.

The categorical score is computed with the help of a hidden layer :
\begin{equation}
c_d(\ket{\sigma})=g\left[\sum_\mu l^\mu_d F\left( \frac{\braket{M^\mu|\sigma}}{T}\right)\right] \label{eq:category}
\end{equation}
$\ket{M^\mu}$s are internal (vectorial) nodes in the the hidden layer, indexed by $\mu$. We use the standard bra-ket $\braket{\cdot|\cdot}$ notation for dot product.   Crucially, because  $\ket{M^\mu}$ and $\ket \sigma$  have the same dimensionality, each $\ket{M^\mu}$ can also be interpreted as a (dual) input, or in the MNIST context, as a picture. Thus $\ket{M^\mu}$s account for a direct representation of what the network learns, which explains why they are usually called 'memories'. $T$ is a fixed temperature which typically 'renormalizes' the dot product around $1$, and we define the rescaled temperature $T_r = \frac{T}{784}$, where $T_r = 1$ corresponds to the self dot product of a picture made only with $\pm 1$ pixels.  
The function $F$ is a rectified Polynomial :
\begin{equation}
F(x)=x^n \quad \mathrm{if } \quad x>0 , \quad 0 \quad \mathrm{otherwise } \label{eq:dfn_F}
\end{equation}
$F$ introduced a new hyperparameter $n$. Intuitively, $n$ is expected to `steepen' the energy landscape around 'true' minima (corresponding to memories in Hopfield's initial picture), thus possibly leading to less spurious minima and more efficient learning/encoding. Indeed, it is now well established that such steepening allows for more 'packing' of information in memory spaces, with an explosion of memory capacity \cite{Demircigil2017}, which also explains the current renewed interest in Hopfied networks.
 $l^\mu_d$ is the label associated to internal memory $\ket{{M}^\mu}$, taking values in the range $[-1,1]$ ($l^\mu_d=1$ means that the memory $\ket{{M}^\mu}$ maximally activates the category $d$) and finally $g$ is a sigmoidal function, taking values between $-1$ and $1$ (typically $\tanh$).
 We refer to Fig. \ref{fig:fig0} A for an illustration of the architecture.
The training consists of optimizing the deviations between $c_d$ and the true categories of the inputs $t_d$. i.e. optimizing by  gradient descent acting on internal memories  $M^\mu$ and labels $l^\mu_d$. More details are given in the Supplement.

All code used for the simulations and for the figures is available at the following repository :
\url{https://github.com/nacer-eb/KrotovHopfieldWaddington}

\section*{Acknowledgement}

We thank members of the Fran\c cois and Mani groups for useful discussions. PF is supported by the Natural Sciences and Engineering Research Council of Canada, Discovery grant program, the Canadian Institutes of Health Research, Project grant program, and the Fonds Courtois. MM was supported by The National 454 Science Foundation-Simons Center for Quantitative Biology at Northwestern University and 455 the Simons Foundation grant 597491. MM is a Simons Investigator. This project has been made possible in part by grant number  DAF2023-329587from the Chan Zuckerberg Initiative DAF, an advised fund of Silicon Valley Community Foundation.

\bibliography{References_Hopfield}

\begin{thebibliography}{10}

\bibitem{Hopfield1982}
JJ Hopfield, Neural networks and physical systems with emergent collective
  computational abilities.
\newblock {\em\protect\JournalTitle{Proc Natl Acad Sci U S A}} \textbf{79},
  2554–2558 (1982).

\bibitem{Ramsauer2020}
H Ramsauer, et~al., Hopfield networks is all you need.
\newblock {\em\protect\JournalTitle{arXiv}}, 2008.02217v1 (2020).

\bibitem{Goodfellow2014}
IJ Goodfellow, O Vinyals, AM Saxe, Qualitatively characterizing neural network
  optimization problems.
\newblock {\em\protect\JournalTitle{arXiv}}, 1412.6544v6 (2014).

\bibitem{Paquette2021}
C Paquette, K Lee, F Pedregosa…, Sgd in the large: Average-case analysis,
  asymptotics, and stepsize criticality.
\newblock {\em\protect\JournalTitle{Proceedings of Thirty Fourth Conference on
  Learning Theory}} (2021).

\bibitem{Rocks2022}
J Rocks, P Mehta, Memorizing without overfitting: Bias, variance, and
  interpolation in overparameterized models.
\newblock {\em\protect\JournalTitle{Phys Rev Res}} \textbf{4}, 013201 (2022).

\bibitem{Tishby}
R Shwartz-Ziv, N Tishby, Opening the black box of deep neural networks via
  information.
\newblock {\em\protect\JournalTitle{arXiv.org}} \textbf{cs.LG} (2017).

\bibitem{Mao2023}
J Mao, et~al., The training process of many deep networks explores the same
  low-dimensional manifold.
\newblock {\em\protect\JournalTitle{arXiv}}, 2305.01604v1 (2023).

\bibitem{Matsushita2022}
Y Matsushita, TS Hatakeyama, K Kaneko, Dynamical systems theory of cellular
  reprogramming.
\newblock {\em\protect\JournalTitle{Physical Review Research}} \textbf{4}
  (2022).

\bibitem{Matsushita2023}
Y Matsushita, K Kaneko, Generic optimization by fast chaotic exploration and
  slow feedback fixation.
\newblock {\em\protect\JournalTitle{Physical Review Research}} \textbf{5}
  (2023).

\bibitem{Rand2021}
D Rand, A Raju, M Sáez, F Corson, E Siggia, Geometry of gene regulatory
  dynamics.
\newblock {\em\protect\JournalTitle{Proc Natl Acad Sci U S A}} \textbf{118},
  e2109729118 (2021).

\bibitem{Seyboldt2022}
R Seyboldt, et~al., Latent space of a small genetic network: Geometry of
  dynamics and information.
\newblock {\em\protect\JournalTitle{Proceedings of the National Academy of
  Sciences}} \textbf{119}, e2113651119 (2022).

\bibitem{Liu2022}
S Liu, P Lemaire, E Munro, M Mani, A mathematical theory for the mechanics of
  three-dimensional cellular aggregates reveals the mechanical atlas for
  ascidian embryogenesis.
\newblock {\em\protect\JournalTitle{bioRxiv}} (2022).

\bibitem{Thura2022}
D Thura, J Cabana, A Feghaly, P Cisek, Integrated neural dynamics of
  sensorimotor decisions and actions.
\newblock {\em\protect\JournalTitle{PLoS biology}} (2022).

\bibitem{Waddington1957}
CH Waddington, {\em The Strategy of the Genes}.
\newblock (Routledge), p. 270 (1957).

\bibitem{Graf2009}
T Graf, T Enver, Forcing cells to change lineages.
\newblock {\em\protect\JournalTitle{Nature}} \textbf{462}, 587–594 (2009).

\bibitem{Lang2014}
A Lang, H Li, J Collins, P Mehta, Epigenetic landscapes explain partially
  reprogrammed cells and identify key reprogramming genes.
\newblock {\em\protect\JournalTitle{PLoS Comput Biol}} \textbf{10}, e1003734
  (2014).

\bibitem{Pusuluri2018}
ST Pusuluri, AH Lang, P Mehta, HE Castillo, Cellular reprogramming dynamics
  follow a simple 1d reaction coordinate.
\newblock {\em\protect\JournalTitle{Physical Biology}} \textbf{15}, 016001
  (2018).

\bibitem{Saez2021}
M Sáez, et~al., Statistically derived geometrical landscapes capture
  principles of decision-making dynamics during cell fate transitions.
\newblock {\em\protect\JournalTitle{Cell Syst}}, S2405–4712(21)00336 (2021).

\bibitem{Husain2020}
K Husain, A Murugan, Physical constraints on epistasis.
\newblock {\em\protect\JournalTitle{Molecular Biology and Evolution}}
  \textbf{37}, 2865–2874 (2020).

\bibitem{Furusawa2018}
C Furusawa, K Kaneko, Formation of dominant mode by evolution in biological
  systems.
\newblock {\em\protect\JournalTitle{Physical Review E}} \textbf{97} (2018).

\bibitem{Gould1979}
SJ Gould, RC Lewontin, The spandrels of san marco and the panglossian paradigm:
  a critique of the adaptationist programme.
\newblock {\em\protect\JournalTitle{Proceedings of the Royal Society of London.
  Series B, Containing Papers of a Biological Character}} \textbf{205},
  581–598 (1979).

\bibitem{Hoover2023}
B Hoover, et~al., Memory in plain sight: A survey of the uncanny resemblances
  between diffusion models and associative memories.
\newblock {\em\protect\JournalTitle{arXiv}}, 2309.16750v1 (2023).

\bibitem{Krotov2016}
D Krotov, JJ Hopfield, Dense associative memory for pattern recognition.
\newblock {\em\protect\JournalTitle{Advances in neural information processing systems}}, (2016).

\bibitem{Krotov2017}
D Krotov, JJ Hopfield, Dense associative memory is robust to adversarial
  inputs.
\newblock {\em\protect\JournalTitle{Advances in neural information processing systems}}, (2017).

\bibitem{Demircigil2017}
M Demircigil, J Heusel, M Löwe, S Upgang, F Vermet, On a model of associative
  memory with huge storage capacity.
\newblock {\em\protect\JournalTitle{Journal of Statistical Physics}}
  \textbf{168}, 288–299 (2017).

\bibitem{lecun1998mnist}
Y LeCun, The mnist database of handwritten digits.
\newblock {\em\protect\JournalTitle{http://yann. lecun. com/exdb/mnist/}}
  (1998).

\bibitem{Goldenfeld2011}
N Goldenfeld, C Woese, Life is physics: Evolution as a collective phenomenon
  far from equilibrium.
\newblock {\em\protect\JournalTitle{Annual Review of Condensed Matter Physics}}
  \textbf{2}, 375–399 (2011).

\bibitem{Freedman2021}
SL Freedman, B Xu, S Goyal, M Mani, A dynamical systems treatment of
  transcriptomic trajectories in hematopoiesis.
\newblock {\em\protect\JournalTitle{Development}}, dev. 201280 (2023).

\bibitem{Yampolskaya2023}
M Yampolskaya, M Herriges, L Ikonomou, D Kotton, P Mehta, sctop:
  physics-inspired order parameters for cellular identification and
  visualization.
\newblock {\em\protect\JournalTitle{bioRxiv}}, 2023.01. 25.525581 (2023).

\bibitem{McInnes2018}
L McInnes, J Healy, J Melville, Umap: Uniform manifold approximation and
  projection for dimension reduction.
\newblock {\em\protect\JournalTitle{arXiv preprint arXiv:1802.03426}} (2018).

\bibitem{Johnson2022}
EM Johnson, W Kath, M Mani, Embedr: distinguishing signal from noise in
  single-cell omics data.
\newblock {\em\protect\JournalTitle{Patterns}} \textbf{3} (2022).

\bibitem{Patcher2023}
T Chari, L Pachter, The specious art of single-cell genomics.
\newblock {\em\protect\JournalTitle{{PLOS} Computational Biology}} \textbf{19},
  e1011288 (2023).

\bibitem{Thom}
Thom, Topological models in biology.
\newblock {\em\protect\JournalTitle{Topology}} (1968).

\bibitem{Francois2023}
P François, New wave theory.
\newblock {\em\protect\JournalTitle{Development}} \textbf{150}, dev201647
  (2023).

\end{thebibliography}

\end{document}